\documentclass[letter]{jpsj3}

\def\runauthor{Yositake {\sc Takane}}
\title{%
Josephson Current through a Planar Junction of Graphene
}

\author{%
Yositake {\sc Takane} and Ken-Ichiro {\sc Imura}
}

\inst{%
Department of Quantum Matter, Graduate School of Advanced Sciences of Matter,\\
Hiroshima University, Higashihiroshima, Hiroshima 739-8530, Japan
}

\recdate{ \hspace{50mm} }

\abst{%
Josephson effect in a planar graphene junction is studied
by assuming that the coupling of a graphene sheet and
two superconductors deposited on its top is described
by a tunneling Hamiltonian.
This model properly takes account of the proximity effect
characteristic to a planar junction,
and allows us to treat monolayer and bilayer cases in a parallel manner.
Applying a quasiclassical Green's function approach to it
we analyze the Josephson critical current $I_{\rm c}$
in a short-junction limit.
As a characteristic feature of the planar junction
we find that $I_{\rm c}$ is a concave function of temperature
at the strong coupling limit while it crosses over to a convex function
with decreasing the coupling strength.
We also find different chemical-potential dependences of $I_{\rm c}$
in the monolayer and bilayer cases.
}

\kword{%
Josephson effect, monolayer graphene, bilayer graphene,
quasiclassical Green's function
}

\begin{document}
\sloppy
\maketitle

Since the realization of a monolayer sheet of graphene,~\cite{novoselov}
extensive studies have been devoted to uncovering unusual electronic
properties of this material.~\cite{castro_neto}
Josephson effect in graphene has been a target of intense
theoretical~\cite{wakabayashi,titov,moghaddam,gonzalez,black-schaffer,hayashi,
hagymasi} and experimental~\cite{heersche,du,ojeda,tomori} studies
during the last few years.
The main interest is focused on how the Josephson current is affected by
the unique band structure of graphene, i.e., the conduction and valence bands
touch conically at $K_{+}$ and $K_{-}$ points in the Brillouin zone,
and the density of states vanishes at the energy
of the band touching point (i.e., Dirac point),
which is set as $\epsilon = 0$ hereafter.
Titov and Beenakker~\cite{titov} calculated the Josephson current through
a monolayer graphene sheet on which two superconducting electrodes are
deposited with separation $L$, under the assumption that carriers are
heavily doped in the region covered by the superconductors.
In the short-junction limit where $L$ is much shorter than
the superconducting coherence length $\xi$,
they obtained the critical current $I_{\rm c}$ at zero temperature
as a function of the chemical potential $\mu$.
It is shown that $I_{\rm c}$ is finite even at $\mu = 0$
and linearly increases with increasing $\mu$.
An experimental result consistent with this prediction has been
reported.~\cite{heersche}

We focus on another interesting aspect of the Josephson effect
in graphene, stemming from the fact that graphene is a unique realization
of an isolated ideal two-dimensional (2D) electron system.
Graphene in a Josephson junction acquires a 2D (planar) contact
with a superconductor, since a natural way to create
a superconductor-graphene-superconductor junction is
to deposit superconducting electrodes on top of a graphene
flake.~\cite{heersche,du,ojeda,tomori}
This is quite contrasting to the case of a usual 2D electron gas
imbedded in a semiconductor hetero-structure that has
a one-dimensional (linear) contact with superconducting electrodes.
However, the previous theoretical studies have not paid attention
to the structure of such a planar junction.
In the model used so far,~\cite{titov,moghaddam,black-schaffer,
hagymasi} an energy-independent effective pair potential $\Delta_{\rm G}$
is induced inside the graphene sheet over the region covered
by the superconductors.
This assumption reduces the planar junction
to a conventional linear junction model.
It is questionable whether the superconducting proximity
effect in a planar junction is fully described
by a conventional model with $\Delta_{\rm G}$.

In this letter we study the stationary Josephson effect in the planar junction
of graphene by employing a simple model in which a graphene sheet is
coupled with superconductors by a tunneling Hamiltonian.~\cite{mcmillan}
This model properly describes the proximity effect characteristic to
a planar junction, which is ignored in the previous theoretical studies,
allowing us to treat monolayer and bilayer cases in a parallel manner.
We apply a quasiclassical Green's function approach~\cite{svidzinskii,kulik}
to our model under an effective mass approximation.~\cite{slonczewski}
When $\mu$ is away from the Dirac point, our approach enables us to derive
a general expression for the Josephson current,
which is applicable for an arbitrary coupling strength $\Gamma$
between the graphene sheet and the superconductors.
Using this expression
we calculate the Josephson current in the short-junction limit.
We show that $I_{\rm c}$ is a concave function of temperature $T$
at the large-$\Gamma$ limit,
while it crosses over to a convex function with decreasing $\Gamma$.
This convex $T$-dependence was not observed
in the previous study~\cite{hagymasi}
based on the model with an energy-independent effective pair potential,
and should be regarded as a characteristic feature of the planar junction.
We also show that the $\mu$-dependence of $I_{\rm c}$ at $T = 0$
qualitatively differs in the monolayer and bilayer cases.
We set $k_{\rm B} = \hbar = 1$ throughout this letter.

\begin{figure}[btp]
\begin{center}
\includegraphics[height=4.5cm]{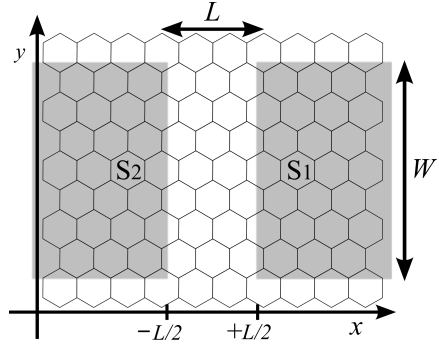}
\end{center}
\caption{Planar junction geometry: Josephson junction consisting of a graphene
sheet on which two superconductors $S_{1}$ and $S_{2}$ of width $W$
are deposited with separation $L$.
The pair potential is assumed to be $\Delta {\rm e}^{{\rm i}\varphi/2}$
in $S_{1}$ and $\Delta {\rm e}^{-{\rm i}\varphi/2}$ in $S_{2}$.
}
\end{figure}
Let us consider a clean graphene sheet on which two superconductors,
${\rm S}_{1}$ and ${\rm S}_{2}$, of width $W$ are deposited
with separation $L$, where ${\rm S}_{1}$ and ${\rm S}_{2}$ occupy
the region of $L/2 \le x$ and that of $x \le -L/2$,
respectively (see Fig.~1).
Note that only the top layer is in contact with
${\rm S}_{1}$ and ${\rm S}_{2}$ in the bilayer case.
Under the condition of $W \gg L$, we regard our system
as being translationally invariant in the $y$-direction.
We assume that carrier doping is uniform in graphene
and that the pair potential $\Delta (x)$ for the superconductors is given by
$\Delta (x) = \Delta {\rm e}^{{\rm i}\varphi/2}$ in ${\rm S}_{1}$ and
$\Delta {\rm e}^{-{\rm i}\varphi/2}$ in ${\rm S}_{2}$.
We use a tunneling Hamiltonian to describe
the coupling of the graphene sheet and the superconductors.
The resulting proximity effect on quasiparticles in graphene
is described by a self-energy~\cite{mcmillan}
for the thermal Green's function given below.
We consider only quasiparticle states near the $K_{+}$ point
because the $K_{+}$ and $K_{-}$ points are degenerate.
To describe quasiparticle states in graphene on the basis of
a tight-binding model,~\cite{slonczewski,wallace}
we introduce nearest-neighbor transfer integral $\gamma_{0}$.
In addition, we employ nearest-neighbor interlayer coupling $\gamma_{1}$
and next nearest-neighbor interlayer coupling $\gamma_{3}$
in the bilayer case.
They are estimated as $\gamma_{0} \approx 2.8$ eV, $\gamma_{1} \approx 0.4$ eV,
and $\gamma_{3} \approx 0.3$ eV.~\cite{castro_neto}
Let us introduce the thermal Green's function
$\check{G}_{j}(\mib{r},\mib{r}';\omega_{n})$
with the Matsubara frequency $\omega_{n} = (2n+1)\pi T$.
The subscript $j = 1$ ($j = 2$) specifies the monolayer (bilayer) case.
Within an effective mass approximation, the Green's function obeys
\begin{align}
      \label{eq:eq-G_j}
  \left( {\rm i}\omega_{n} \check{\tau}_{4\times4}^{z}-\check{H}_{j}
          -\check{\Sigma}_{j}
  \right) \check{G}_{j}(\mib{r},\mib{r}';\omega_{n})
  = \check{\tau}_{4\times4}^{0}\delta(\mib{r}-\mib{r}')
\end{align}
with $\check{\tau}_{4\times4}^{z} = {\rm diag}(1,1,-1,-1)$,
$\check{\tau}_{4\times4}^{0}={\rm diag}(1,1,1,1)$,
$\check{H}_{j} = {\rm diag} \left( H_{j}, H_{j} \right)$.
The $2 \times 2$ effective Hamiltonian $H_{j}$
for low-energy quasiparticles is given by
\begin{align}
      \label{eq:H1}
  H_{1}   & = \left( \begin{array}{cc}
                       -\mu & \gamma \hat{k}_{-} \\
                       \gamma \hat{k}_{+} & -\mu
                     \end{array}
              \right)
\end{align}
for the monolayer case~\cite{slonczewski}
with $\gamma = (\sqrt{3}/2)\gamma_{0}a$ ($a$: lattice constant)
and $\hat{k}_{\pm} = -{\rm i}\partial_{x}\pm\partial_{y}$, and
\begin{align}
      \label{eq:H2}
  H_{2}
  = \left( \begin{array}{cc}
             -\mu & - \alpha \hat{k}_{+}^{2}-\beta \hat{k}_{-} \\
             - \alpha \hat{k}_{-}^{2}-\beta \hat{k}_{+} & -\mu
           \end{array}
    \right)
\end{align}
for the bilayer case~\cite{maccann}
with $\alpha = \gamma^{2}/\gamma_{1}$ and
$\beta = (\sqrt{3}/2)\gamma_{3}a$ characterizing the trigonal warping.
The self-energy $\check{\Sigma}_{j}$ is  given by~\cite{mcmillan}
\begin{align}
     \label{eq:self-e}
 \check{\Sigma}_{j}
      = \frac{-{\rm i}\Gamma}{\sqrt{\Delta^{2}+\omega_{n}^{2}}}
        \left( \begin{array}{cc}
                  \omega_{n} \chi_{2 \times 2}^{(j)}
                    & \Delta(x) \chi_{2 \times 2}^{(j)} \\
                  \Delta(x)^{*} \chi_{2 \times 2}^{(j)}
                    & -\omega_{n} \chi_{2 \times 2}^{(j)}
               \end{array}
        \right)
        \theta(|x|-L/2) ,
\end{align}
where $\Gamma$ characterizes the coupling strength between
the graphene sheet and the superconductors,
$\chi_{2 \times 2}^{(1)} = {\rm diag}(1,1)$,
and $\chi_{2 \times 2}^{(2)} = {\rm diag}(1,0)$.
The matrix form of $\chi_{2 \times 2}^{(2)}$ reflects the fact
that only the top layer is in contact with the superconductors
in the bilayer case.~\cite{takane}
The off-diagonal elements of $\check{\Sigma}_{j}$ are regarded as
an energy-dependent effective pair potential,
while the diagonal elements describe
renormalization of a quasiparticle energy.
If the $\omega_{n}$-dependence is ignored by setting $\omega_{n} = 0$,
our model is reduced to the conventional one.~\cite{titov,moghaddam,hagymasi}

Hereafter we restrict our attention to the moderate doping regime
of $\gamma_{0}, \gamma_{1} \gg \mu \gg \Delta$.
To introduce the quasiclassical Green's function,
we perform a Fourier transformation as
\begin{align}
  \check{G}_{j}(\mib{p},\mib{r};\omega_{n})
 = \int {\rm d}^{2}s {\rm e}^{-{\rm i}\mib{p}\cdot{\mib{s}}}
   \check{G}_{j}
   \left(\mib{r}+\frac{\mib s}{2},\mib{r}-\frac{\mib s}{2};\omega_{n}
   \right) ,
\end{align}
where ${\mib p} = (p_{x},p_{y})$.
In the monolayer case of $j = 1$, the Green's function obeys
\begin{align}
  \left( {\rm i}\omega_{n} \check{\tau}_{4\times4}^{z}-\check{\mathcal{H}}_{1}
          -\check{\Sigma}_{1}
  \right) \check{G}_{1}(\mib{p},\mib{r};\omega_{n})
  = \check{\tau}_{4\times4}^{0} ,
\end{align}
where the $4 \times 4$ Hamiltonian $\check{\mathcal{H}}_{1}$ is given by
$\check{\mathcal{H}}_{1} = {\rm diag} ( \tilde{H}_{1}+h_{1},
\tilde{H}_{1}+h_{1} )$ with
\begin{align}
 \tilde{H}_{1}
     & = \left( \begin{array}{cc}
                  -\mu & \gamma (p_{x}-{\rm i}p_{y}) \\
                   \gamma (p_{x}+{\rm i}p_{y}) & -\mu
                \end{array}
         \right) ,
         \\
 h_{1}
     & = \left( \begin{array}{cc}
                   0 & \frac{\gamma}{2} \hat{k}_{-} \\
                    \frac{\gamma}{2} \hat{k}_{+} & 0
                \end{array}
         \right) .
\end{align}
The $2\times2$ matrix $\tilde{H}_{1}$ is diagonalized as
$u_{\mib p}^{\dagger} \tilde{H}_{1} u_{\mib p}
= {\rm diag} \left( \gamma p -\mu, -\gamma p -\mu \right)$ in terms of
\begin{align}
  u_{\mib p}
  = \frac{1}{\sqrt{2}}
    \left( \begin{array}{cc}
             1 & - {\rm e}^{-{\rm i}\phi_{\mib p}} \\
             {\rm e}^{{\rm i}\phi_{\mib p}} & 1
           \end{array}
    \right) ,
\end{align}
where $p = |{\mib p}|$ and
$\phi_{\mib{p}} = {\rm arg}\{ p_{x}+{\rm i}p_{y}\}$.
As long as $\mu \gg \Delta$, the subband with the energy dispersion
$-\gamma p -\mu$ is irrelevant in the superconducting proximity effect.
Therefore, we are allowed to consider only the relevant subband with
the energy dispersion $\gamma p -\mu$.
In accordance with this observation, we transform $\check{G}_{1}$ as
$\check{\mathcal{G}}_{1}(\mib{p},\mib{r};\omega_{n}) =
\check{U}_{\mib p}^{\dagger} \check{G}_{1}(\mib{p},\mib{r};\omega_{n})
\check{U}_{\mib p}$
with $\check{U}_{\mib p} = {\rm diag} \left( u_{\mib p}, u_{\mib p} \right)$
and retain only
the $(1,1)$-, $(1,3)$-, $(3,1)$-, and $(3,3)$-elements.~\cite{takane}
Accordingly, we define $G_{1}$ as
\begin{align}
 G_{1}
 = \left( \begin{array}{cc}
             \bigl[\check{\mathcal{G}}_{1}\bigr]_{1,1}
             & \bigl[\check{\mathcal{G}}_{1}\bigr]_{1,3} \\
             \bigl[\check{\mathcal{G}}_{1}\bigr]_{3,1}
             & \bigl[\check{\mathcal{G}}_{1}\bigr]_{3,3}
          \end{array}
   \right) .
\end{align}
The Green's function $G_{1}$ approximately satisfies
\begin{align}
      \label{eq:qc-GN0}
  \left[ {\rm i}\omega_{n} \tau_{2\times2}^{z}
         - \left(\gamma p -\mu + \frac{\rm i}{2}{\mib v}_{1}({\mib p})
                 \cdot \nabla
           \right)
         - \Sigma_{1}
  \right] G_{1}(\mib{p},\mib{r};\omega_{n})
  = \tau_{2\times2}^{0} ,
\end{align}
where $\tau_{2\times2}^{z} = {\rm diag}(1,-1)$,
$\tau_{2\times2}^{0} = {\rm diag}(1,1)$,
the velocity ${\mib v}_{1}({\mib p})$ is given by
${\mib v}_{1}({\mib p}) = \gamma (\cos\phi_{\mib p}, \sin\phi_{\mib p})$, and
\begin{align}
 \Sigma_{1}
   = \frac{-{\rm i}\Gamma}{\sqrt{\Delta^{2}+\omega_{n}^{2}}}
     \left( \begin{array}{cc}
               \omega_{n} & \Delta(x) \\
               \Delta(x)^{*} & -\omega_{n}
            \end{array}
     \right)
     \theta(|x|-L/2) .
\end{align}
By repeating the argument similar to this we can show that
the Green's function for the bilayer case satisfies
\begin{align}
      \label{eq:qc-GNm}
  \left[ {\rm i}\omega_{n} \tau_{2\times2}^{z}
         - \left(\epsilon_{\mib p}-\mu
         + \frac{\rm i}{2}{\mib v}_{2}({\mib p})\cdot \nabla
           \right)
         - \Sigma_{2}
  \right] G_{2}(\mib{p},\mib{r};\omega_{n})
  = \tau_{2\times2}^{0} ,
\end{align}
where $\epsilon_{\mib p}=[(\alpha p^{2})^{2}
+2\alpha\beta p^{3}\cos(3\phi_{\mib p})+(\beta p)^{2}]^{1/2}$,
and $\Sigma_{2} = \Sigma_{1}/2$.
The $x$- and $y$-components of ${\mib v}_{2}({\mib p})$ are given by
\begin{align}
  v_{2x}({\mib p})
  & = 2\alpha p \cos(\theta_{\mib p}-\phi_{\mib p}) 
      + \beta \cos \theta_{\mib p} ,
            \\
  v_{2y}({\mib p})
  & = 2\alpha p \sin(\theta_{\mib p}-\phi_{\mib p}) 
      - \beta \sin \theta_{\mib p} ,
\end{align}
where $\theta_{\mib p} = {\rm arg}\{
\alpha p^{2}{\rm e}^{{\rm i}2\phi_{\mib p}}
+ \beta p{\rm e}^{-{\rm i}\phi_{\mib p}} \}$.
Note that $\Sigma_{2}$ is smaller by a factor of two than $\Sigma_{1}$
reflecting the fact that only the top layer is in contact
with the superconductors.

We define the quasiclassical Green's function
$G_{j}(\mib{n},\mib{r};\omega_{n})$ with ${\mib n}={\mib p}/p$
as~\cite{eilenberger,larkin}
\begin{align}
       \label{eq:def_qcG}
  G_{j}(\mib{n},\mib{r};\omega_{n})
  = \frac{\rm i}{\pi}\int {\rm d}\xi_{\mib p}
    G_{j}(\mib{p},\mib{r};\omega_{n}) ,
\end{align}
where a diverging contribution must be subtracted,
and $\xi_{\mib p} = \gamma p - \mu$ for $j = 1$
and $\xi_{\mib p} = \epsilon_{\mib p} - \mu$ for $j = 2$.
Applying a standard procedure~\cite{larkin}
to eqs.~(\ref{eq:qc-GN0}) and (\ref{eq:qc-GNm}),
we can show that the quasiclassical Green's function satisfies
\begin{align}
      \label{eq:eilenberger_1}
  {\rm i}\omega_{n} \left[\tau_{2\times2}^{z}, G_{j}\right]
  + {\rm i}{\mib v}_{{\rm F}j}(\mib{n})\cdot \nabla G_{j}
  - \left[\Sigma_{j}, G_{j}\right] = 0 ,
\end{align}
where ${\mib v}_{{\rm F}j}(\mib{n})$ represents the Fermi velocity
in the momentum direction denoted by $\mib n$.
We express the elements of $G_{j}(\mib{n},\mib{r};\omega_{n})$ as
$[G_{j}]_{1,1}=-[G_{j}]_{2,2}=g_{j}$, $[G_{j}]_{1,2}=f_{j}$,
and $[G_{j}]_{2,1}=f^{\dagger}_{j}$.
Equation~(\ref{eq:eilenberger_1}) yields
\begin{align}
       \label{eq:eilenberger_21}
  & {\mib v}_{{\rm F}j}({\mib n})\cdot \nabla g_{j}
    = \zeta_{j}(x,\omega_{n})
      \left[\Delta(x)^{*}f_{j} - \Delta(x)f_{j}^{\dagger} \right] ,
           \\
       \label{eq:eilenberger_22}
  &   2 \left[1 + \zeta_{j}(x,\omega_{n})\right]\omega_{n} f_{j}
    + {\mib v}_{{\rm F}j}({\mib n})\cdot \nabla f_{j}
    = 2\zeta_{j}(x,\omega_{n})\Delta(x) g_{j} ,
           \\
       \label{eq:eilenberger_23}
  &   2 \left[1 + \zeta_{j}(x,\omega_{n})\right]\omega_{n} f_{j}^{\dagger}
    - {\mib v}_{{\rm F}j}({\mib n})\cdot \nabla f_{j}^{\dagger}
    = 2\zeta_{j}(x,\omega_{n})\Delta(x)^{*} g_{j} ,
\end{align}
where
\begin{align}
   \zeta_{j}(x,\omega_{n})
   = \frac{\Gamma_{j}}{\sqrt{\Delta^{2}+\omega_{n}^{2}}}
     \theta(|x|-L/2)
\end{align}
with $\Gamma_{1} = \Gamma$ and $\Gamma_{2} = \Gamma/2$.
Note that although we are considering a Josephson junction of graphene,
only the Fermi velocity reflects a feature of graphene
in eqs.~(\ref{eq:eilenberger_21})-(\ref{eq:eilenberger_23}).
We point out that these equations are also applicable to a conventional
planar junction of a 2D electron system.~\cite{volkov}

We derive a general expression for the dc Josephson current
$I_{j}(\varphi)$ in terms of the quasiclassical Green's function,
where $j = 1$ and $2$ correspond to the monolayer and bilayer cases,
respectively.
We ignore the trigonal warping effect by setting $\beta = 0$.
With this simplification, the Fermi velocity is expressed as
${\mib v}_{{\rm F}j} = v_{{\rm F}j}(\cos\phi,\sin\phi)$ with
$v_{{\rm F}1} = \gamma$ and
$v_{{\rm F}2} = 2\gamma \sqrt{\mu/\gamma_{1}}$.
The density of states per spin at the Fermi level is given by
$N_{1}(0) = \mu/(\pi \gamma^{2})$ and
$N_{2}(0) = \gamma_{1}/(2\pi\gamma^{2})$,
in which the two-fold valley degeneracy is included.
The Josephson current is expressed as~\cite{svidzinskii,kulik}
\begin{align}
  I_{j}(\varphi)
  = 2\pi e N_{j}(0) W \int_{-\frac{\pi}{2}}^{\frac{\pi}{2}}
    \frac{{\rm d}\phi}{\pi} v_{{\rm F}jx}
    T \sum_{\omega_{n}} {\rm Im}\left\{g_{j}(\mib{n},x;\omega_{n})\right\} .
\end{align}
Note that in the ${\mib r}$-dependence of $g_{j}$, only $x$ is relevant
since $W \gg L$ is assumed.
Because $I_{j}(\varphi)$ is independent of $x$ for $|x| \le L/2$,
we evaluate it at $x = L/2$.
Parameterizing $x$ as $x = (v_{{\rm F}j}\cos\phi)t$,
we solve eqs.~(\ref{eq:eilenberger_21})-(\ref{eq:eilenberger_23})
with $g_{j}^{2} + f_{j}f_{j}^{\dagger} = 1$
under the boundary condition of
\begin{align}
  & \lim_{t \to \pm \infty}g_{j}(\mib{n},t;\omega_{n})
  = \frac{\tilde{\omega}}{\tilde{\Omega}} ,
     \\
  & \lim_{t \to \pm \infty}f_{j}(\mib{n},t;\omega_{n})
  = \frac{\tilde{\Delta} {\rm e}^{\pm {\rm i}\varphi/2}}
         {\tilde{\Omega}} ,
\end{align}
where
$\tilde{\omega} = \left[1 + \zeta_{j}(x,\omega_{n})\right]\omega_{n}$,
$\tilde{\Delta} = \zeta_{j}(x,\omega_{n}) \Delta$, and
$\tilde{\Omega} = \sqrt{\tilde{\Delta}^{2}+\tilde{\omega}^{2}}$.
Note that $x = L/2$ corresponds to
$t_{+} \equiv L/(2v_{{\rm F}j}\cos\phi)$.
A straightforward calculation yields
\begin{align}
 g_{j}(\mib{n},t_{+};\omega_{n})
 = \frac{\tilde{\omega}}{\tilde{\Omega}}
   + \frac{ \tilde{\Delta}^{2}
            \left( \sinh\kappa \cos\frac{\varphi}{2}
                 + {\rm i}
                   \cosh\kappa \sin\frac{\varphi}{2}
            \right) }
          { \tilde{\Omega}
            \left( \tilde{\Omega} \cosh\kappa
                 + \tilde{\omega} \sinh\kappa \right) \cos\frac{\varphi}{2}
          + {\rm i}\tilde{\Omega}
            \left( \tilde{\Omega} \sinh\kappa
                 + \tilde{\omega} \cosh\kappa \right) \sin\frac{\varphi}{2}
          }
\end{align}
with $\kappa = \omega_{n} L/(v_{{\rm F}j}\cos\phi)$.
We obtain
\begin{align}
       \label{eq:I-nm-1}
 I_{j}(\varphi)
  = 2\pi e N_{j}(0) W
    \int_{-\frac{\pi}{2}}^{\frac{\pi}{2}}
    \frac{{\rm d}\phi}{\pi} v_{{\rm F}jx}
    T \sum_{\omega_{n}}
    \frac{ \tilde{\Delta}^{2} \sin\varphi }
         { \left( \tilde{\Omega}^{2}+\tilde{\omega}^{2} \right)
           \cosh2\kappa
         + 2\tilde{\omega}\tilde{\Omega}\sinh2\kappa
         + \tilde{\Delta}^{2} \cos\varphi } .
\end{align}
Using this general expression one can numerically calculate
the Josephson current in the planar junction for arbitrary parameters.
In the strong-coupling limit of $\Gamma_{j} \gg \Delta_{0}$
with $\Delta_{0}$ being the magnitude of the pair potential at $T = 0$,
eq.~(\ref{eq:I-nm-1}) is reduced to the ordinary expression for
the Josephson current through a 2D electron gas of finite area
placed between two superconductors.~\cite{svidzinskii,furusaki}

Below we focus on the short-junction limit of
$L \ll \xi$, where $\xi \equiv v_{{\rm F}j}/(2\pi \Delta_{0})$
is the superconducting coherence length,
and study the behavior of the critical current for an arbitrary $\Gamma_{j}$.
We can approximate as $\cosh2\kappa \approx 1$ and $\sinh2\kappa \approx 0$
in the short-junction limit, and obtain
\begin{align}
       \label{eq:I-nm-2}
 I_{j}(\varphi)
   = e \mathcal{N}_{j} T \sum_{\omega_{n}}
     \frac{ \Gamma_{j}^{2}\Delta^{2} \sin\varphi }
          { \omega_{n}^{2}
            \left( \Delta^{2}+\omega_{n}^{2}
                   + 2\Gamma_{j}\sqrt{\Delta^{2}+\omega_{n}^{2}}
                   + \Gamma_{j}^{2} \right)
          + \Gamma_{j}^{2}\Delta^{2} \cos^{2}\frac{\varphi}{2} } ,
\end{align}
where $\mathcal{N}_{j} \equiv 2v_{{\rm F}j} N_{j}(0) W$
represents the number of conducting channels.
Before considering the critical current,
let us observe the behavior of $I_{j}(\varphi)$
in the strong and weak coupling limits.
Firstly we consider the strong coupling limit of
$\Gamma_{j} \gg \Delta_{0} \ge \Delta$.
In this case we can ignore all terms in the parentheses
except for $\Gamma_{j}^{2}$, and obtain
\begin{align}
       \label{eq:I-nm-KO2}
 I_{j}(\varphi)
   = e\mathcal{N}_{j}\Delta
     \sin\frac{\varphi}{2}
     \tanh\left(\frac{\Delta \cos\frac{\varphi}{2}}{2T}\right) ,
\end{align}
This is identical to the result derived
by Kulik and Omel'yanchek (KO).~\cite{kulik}
We next consider the weak coupling limit of $\Delta_{0} \gg \Gamma_{j}$.
In the low-temperature regime of $\Delta \gg \Gamma_{j} \gg T$,
we can ignore all terms in the parentheses
except for $\Delta^{2}$, and obtain~\cite{volkov}
\begin{align}
       \label{eq:I-nm-volkov}
 I_{j}(\varphi)
   = e\mathcal{N}_{j}\Gamma_{j}
     \sin\frac{\varphi}{2}
     \tanh\left(\frac{\Gamma_{j} \cos\frac{\varphi}{2}}{2T}\right) .
\end{align}
When $T_{\rm c} \gtrsim T$,
there holds $T \gg \Gamma_{j}, \Delta$ which enables us
to retain only $\omega_{n}^{4}$ in the denominator.
We thus obtain
\begin{align}
       \label{eq:I-nm-Tc}
 I_{j}(\varphi)
   = e\mathcal{N}_{j}
     \frac{\Gamma_{j}^{2}\Delta^{2}}{48T^{3}}\sin\varphi .
\end{align}
This $T$-dependence is qualitatively different from
the KO result which yields
$I(\varphi)=e\mathcal{N}\Delta^{2}(4T)^{-1}\sin\varphi$ near $T_{\rm c}$.

To observe the $T$-dependence of the critical current
$I_{{\rm c}}^{(j)}\equiv{\rm max}_{\varphi} \left\{I_{j}(\varphi)\right\}$,
we numerically calculate $I_{{\rm c}}^{(j)}$
for several values of $r \equiv \Gamma_{j}/\Delta_{0}$
on the basis of eq.~(\ref{eq:I-nm-2}).
The $T$-dependence of $\Delta$ is determined by the gap equation
$1 = \lambda_{\rm int} 
\int_{0}^{\epsilon_{\rm D}} {\rm d}\epsilon
\tanh(\frac{\sqrt{\epsilon^{2}+\Delta^{2}}}{2T})/
\sqrt{\epsilon^{2}+\Delta^{2}}$,
where $\lambda_{\rm int}$ is the dimensionless interaction constant,
and the Debye energy is chosen as $\epsilon_{\rm D}/\Delta_{0} = 200$.
The critical current $I_{\rm c}^{(j)}$ normalized by
$I_{0}^{(j)} \equiv e\mathcal{N}_{j}\Delta_{0}$ is shown in Fig.~2
as a function of $T/T_{\rm c}$ for $r = 0.2$, $1.0$, $5.0$,
and $r \to \infty$ at which the KO result is reproduced.
Note that $I_{\rm c}^{(j)}/I_{0}^{(j)}$ does not depend on $j$.
We observe that
$I_{\rm c}^{(j)}$ is a concave function of $T$ at $r \to \infty$,
while it crosses over to a convex function with decreasing $r$.
Such a convex $T$-dependence was not observed
in the previous study~\cite{hagymasi}
based on an energy-independent effective pair potential model.
The coupling strength $\Gamma_{j}$ crucially affects
the $T$-dependence of the critical current.
\begin{figure}[btp]
\begin{center}
\includegraphics[height=5.0cm]{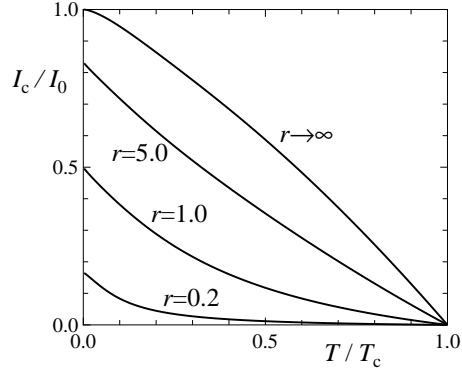}
\end{center}
\caption{Temperature dependence of the normalized critical current
$I_{\rm c}^{(j)}/I_{0}^{(j)}$ for $r = 0.2$, $1.0$, $5.0$,
and $r \to \infty$, where $r \equiv \Gamma_{j}/\Delta_{0}$.
}
\end{figure}

Let us finally evaluate the critical current at $T = 0$.
As an intermediate step, we derive a convenient analytic formula
by approximating the $\omega_{n}$-dependence
of the denominator of eq.~(\ref{eq:I-nm-2})
as $2\Gamma_{j}\sqrt{\Delta^{2}+\omega_{n}^{2}} \approx 2\Gamma_{j}\Delta$.
This is justified either in the low-temperature limit of $T_{\rm c} \gg T$
for an arbitrary $\Gamma_{j}$,
or in the strong and weak coupling limits for an arbitrary $T$.
Performing the summation over $\omega_{n}$ in terms of a contour integral,
we obtain
\begin{align}
       \label{eq:I-nm-result}
 I_{j}(\varphi)
   = e\mathcal{N}_{j}
     \frac{\Gamma_{j}^{2}\Delta^{2} \sin\varphi}
          {\sqrt{E_{+}(\varphi)E_{-}(\varphi)}}
     \left[\frac{\tanh\left(\frac{E_{+}(\varphi)-E_{-}(\varphi)}{4T}\right)}
                {E_{+}(\varphi)-E_{-}(\varphi)}
         - \frac{\tanh\left(\frac{E_{+}(\varphi)+E_{-}(\varphi)}{4T}\right)}
                {E_{+}(\varphi)+E_{-}(\varphi)}
     \right] ,
\end{align}
where $E_{\pm}(\varphi) = [(\Delta+\Gamma_{j})^{2}
\pm 2\Gamma_{j}\Delta\cos(\varphi/2)]^{1/2}$.
One can verify that eq.~(\ref{eq:I-nm-result}) reproduces
the correct asymptotic behaviors, eqs.~(\ref{eq:I-nm-KO2})-(\ref{eq:I-nm-Tc}),
in the corresponding limits.
Note that eq.~(\ref{eq:I-nm-result}) is maximized
at $\varphi = \pi$ (mod $2\pi$) in the limit of $T \to 0$.
The critical current is determined as $I_{\rm c}^{(j)} = e\mathcal{N}_{j}
\Gamma_{j}\Delta_{0}/(\Delta_{0}+\Gamma_{j})$, which yields
\begin{align}
  I_{\rm c}^{(1)}
   = e\frac{2\mu W}{\pi \gamma}
     \frac{\Gamma\Delta_{0}}{\Delta_{0}+\Gamma}
\end{align}
for the monolayer, and
\begin{align}
  I_{\rm c}^{(2)}
   = e\frac{2\mu W}{\pi \gamma}\sqrt{\frac{\gamma_{1}}{\mu}}
     \frac{\frac{\Gamma}{2}\Delta_{0}}{\Delta_{0}+\frac{\Gamma}{2}}
\end{align}
for the bilayer cases.
We see that $I_{\rm c}^{(1)}$ is proportional to $\mu$,
which is consistent with the result reported in ref.~\citen{titov},
while $I_{\rm c}^{(2)}$ is proportional to $\sqrt{\mu}$.
Roughly speaking, the critical current in the bilayer case
is greater than that in the monolayer case by a factor of
$\sqrt{\gamma_{1}/\mu}$.

In summary we have proposed a model for a planar Josephson junction
of graphene, and derived a general expression for the Josephson current
at moderate doping in the quasiclassical Green's function approach.
Much emphasis has been on the behavior of the Josephson current
in the short-junction limit in monolayer and bilayer graphene junctions.
It was demonstrated that the coupling strength crucially affects the
temperature dependence of the critical current in an unexpected manner.
This should be regarded as a characteristic feature of the planar junction.
We have also shown that the chemical-potential dependence of the critical
current qualitatively differs in the monolayer and bilayer cases.
Finally we point out that our argument can be extended
to a multilayer case.~\cite{takane}
Such an extension will be reported elsewhere.

\section*{Acknowledgment}

This work was supported in part by a Grant-in-Aid for Scientific
Research (C) (No. 21540389)
from the Japan Society for the Promotion of Science.

\end{document}